# Equivalence and origin of discrepancies between transfer matrix method and multiple reflection model of surface induced second harmonic generation

Deependra Singh Gaur[1,*], Jakub Lukeš[1,2] and Karel Žídek[1]

[1]Institute of Plasma Physics of the Czech Academy of Sciences, U Slovanky 2525/1a, 182 00 Prague 8, Czech Republic

[2]Technical University of Liberec, Studentská 1402/2, 461 17, Liberec, Czech Republic

*gaur@ipp.cas.cz

**Abstract**

Second harmonic generation (SHG) spectroscopy is a sensitive tool to study surfaces and interfaces in thin films. To understand the layer boundaries and interpret the measured data, it is essential to use a theoretical model describing the SHG generation in the thin film. The transfer matrix method and multiple reflection model are two numerical techniques which have been commonly utilized in the calculation of surface-induced SHG. Here, we demonstrate that, despite the expected equivalence of the approaches, the two methods can lead to controversial results for the same input properties and second-order susceptibility tensors. The discrepancies resulting in different SHG intensity from both methods are caused by inconsistent numerical implementation of the interface SHG. We describe the conditions for the implementation to achieve the equivalence between the two models. Moreover, we identify the physical constraint that has to be imposed in numerical implementation to get identical results from both methods.

## 1. Introduction

SHG is the second-order nonlinear optical phenomenon where two photons with fundamental frequency $\omega$ generates a photon with doubled frequency $2\omega$ [1]. This process is of very high importance in a number of spectroscopic methods. For instance, SHG is a sensitive tool able to track the properties of surfaces and interfaces in thin films.

SHG in thin films mainly originates from two sources, bulk and surface contributions, where the symmetry of the medium plays a prominent role. Within the electronic dipole approximation, bulk SHG is forbidden in centrosymmetric materials due to inversion symmetry. However, multipolar contributions such as electric quadrupole and magnetic dipoles can lead to nonzero bulk SHG. Moreover, inversion symmetry is necessarily broken at the surface, resulting in nonvanishing surface-induced SHG even in a centrosymmetric material [1,2]. Thus, this symmetry-sensitive characteristic enables SHG to be a non-invasive, contactless optical technique for optical metrology, allowing surface and interface monitoring [3-7].

Surface-induced SHG primarily arises from both structural asymmetry and field discontinuity originating from the refractive index contrast. [8,9]. Particularly, refractive index mismatch across the boundary drives rapid change in the normal component of the electric field, prompting nonlocal surface response. Therefore, nonlinear response emerging through the gradient of the field includes multipolar alongside dipole contribution at the surface SHG whose distinction has been a long-standing problem [10]. Nevertheless, two beam method and mode confinement in waveguide allow to separate surface contribution and multipolar bulk contribution in a centrosymmetric material [11]. Surface SHG is not limited to a particular centrosymmetric material, as it has been observed in many material surfaces and nanostructures including metals, semiconductors, chiral surfaces and metasurfaces [12-15]. Surface SHG can be further enhanced by utilizing surface plasmon resonance or symmetry breaking through the light. [16-18]. Moreover, multiple stacks of centrosymmetric thin films naturally leverage with many buried interfaces, which enhance the surface SHG [19].

Several theoretical models have been developed to quantify the surface SHG [20,21]. J. Sipe has developed a new formalism of surface SHG utilizing the Green function, providing phenomenological treatment of surface-induced SHG in multiple geometries [22,23]. Later, Gielis et. al. have also proposed a similar approach, namely a multiple-reflection model to calculate the SHG from surfaces of a thin film [24], but the numerical implementation of these approaches is cumbersome, particularly when number of layers is more than one.

However, the localized optical states observed in multilayer structures significantly influence the nonlinear interaction and resulting multilayer SHG. Therefore, a computationally convenient method called transfer matrix method (TMM) has also been developed in parallel that can be extended for multilayer structures [25]. For a periodic multilayer structure, an analytic expression for the electromagnetic field has been evaluated in the undepleted regime using the matrix method [26]. Interestingly, the Green's function approach can be integrated with TMM for investigating the multilayer structures [27,28]. TMM allows one to study a wide range of nonlinear interactions with fewer physical constraints. Therefore, it emerges as an efficient method for examining the electromagnetic interaction with thin film stacks [29-31]. Recently, TMM based on generalized boundary conditions has been reported for calculating surface SHG from multilayer structures where bulk effects are neglected [32].

TMM and the multiple reflection approach (Gielis et al. implementation of the Green's function) are two popular numerical techniques for calculating surface-induced SHG [32-34]. Both methods were derived from Maxwell equations and therefore should be identical under the same physical conditions. But both methods provide distinct surface/interface SHG for one physical system, and their implementation leads to diverse results. Namely, the surface susceptibilities $\chi^{(2)}$values depend on the conventions used, which raises a concern on theoretical consistency of both frameworks and the comparability of the surface/interface $\chi^{(2)}$ value in the literature. To compare the results between the two methods, it is indispensable to create a solid link between these two approaches.

In this work, we unveil the possible discrepancies leading to the disagreement in the surface-induced SHG calculated by both theoretical models and provide diverse nonlinear surface susceptibilities. For this purpose, we numerically implement both methods and compare them systematically. This comparison allows us to identify the possible reasons responsible for disagreement in the surface SHG, particularly the position of the induced nonlinear polarization sheet in both methods. Moreover, rescaling of nonlinear surface susceptibilities in the multiple reflection model and source matrix in TMM affect the total SHG which causes the distinct total SHG in both methods. These ambiguities in the SHG results can be removed by adopting a consistent numerical implementation of both frameworks. Therefore, we demonstrate the equivalence between both methods under certain conditions which allow to establish the relation between nonlinear surface susceptibilities calculated from different methods.

**2. Theoretical modelling**

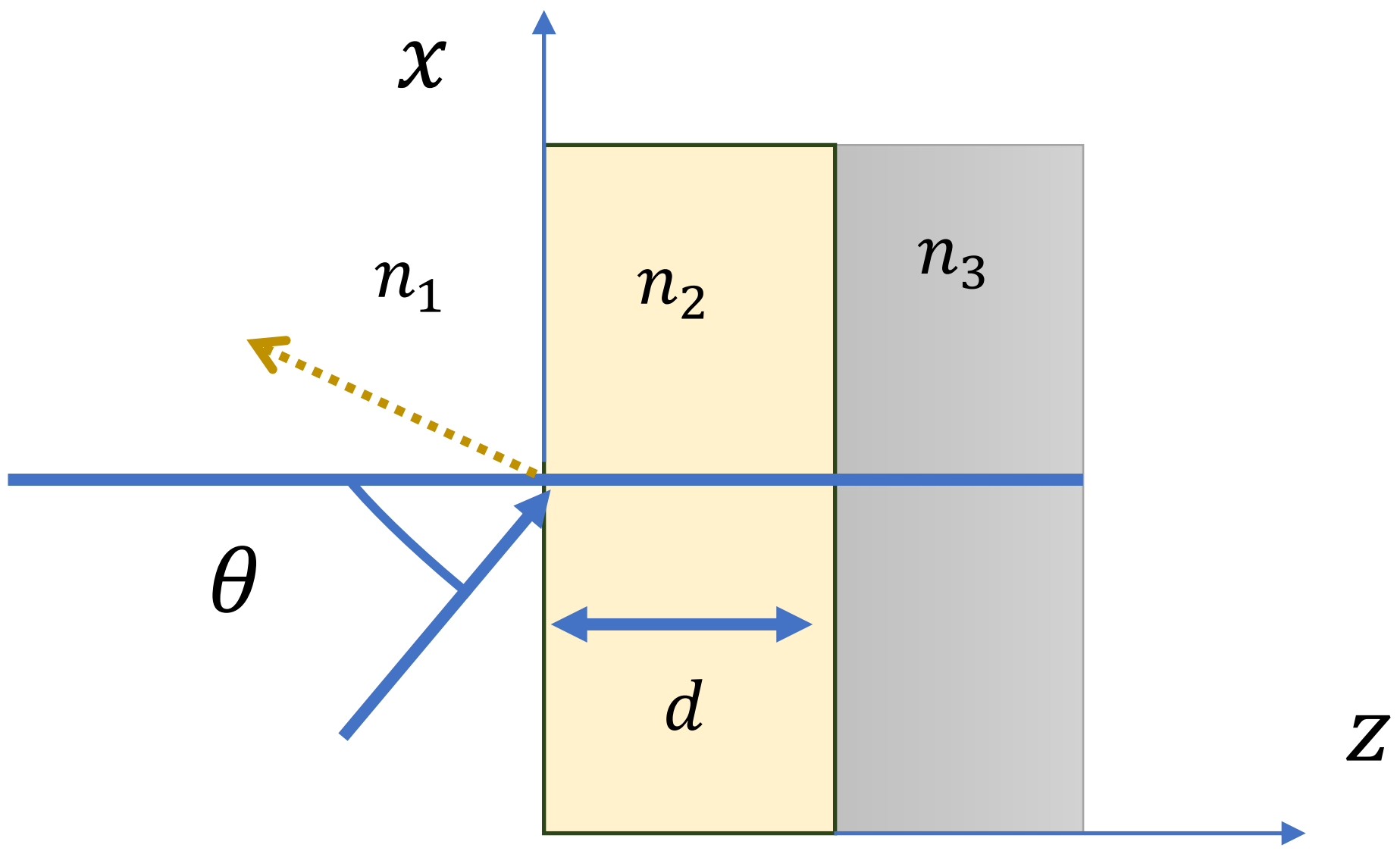

Figure 1: Schematic of thin film structure with thickness $d$ and refractive indices $n_2$ deposited on to the semi-infinite substrate of refractive index $n_3$

We consider a dielectric thin film of a centrosymmetric material deposited on a semi-infinite substrate as illustrated schematically in Fig. 1. The dielectric constants or refractive indices of the superstrate, layer and substrate can be represented as $n_i = \sqrt{\epsilon_i}$ where $i = 1,2,3$, respectively. We denote $d$ as the thickness of the film. Here, $z$ axis is assumed to be perpendicular to thin film interface and $xy$ plane coincides with the interface plane. The incident plane wave with fundamental frequency $\omega$ propagates from left to right (positive $z$ direction) at incident angle $\theta$.

In this system, the bulk electric dipole contribution is assumed to vanish due to inversion symmetry and the multipolar contribution is not considered in the theory derivation. Therefore, only surface/interface induced SHG is the dominant contribution which can be calculated employing many approaches. In this work, we focus on the multiple reflection model and transfer matrix method [24,32]. We discuss these two methods in brief and examine their theoretical consistency using numerical simulation.

### 2.1 Multiple reflection model

We will follow here the derivation provided by Gielis et. al in ref. 24, using the same notation. For the full derivation and reasoning, we point the reader to this article.

An optical field propagating inside thin film encounters multiple reflections. When optical field propagates from medium $i$ to medium $j$, reflection and transmission can be described by the Fresnel coefficients,

$$r_{ij}^{s} = \frac{k_{iz} - k_{jz}}{k_{iz} + k_{jz}}, \qquad r_{ij}^{p} = \frac{k_{iz}\epsilon_j - k_{jz}\epsilon_i}{k_{iz}\epsilon_j + k_{jz}\epsilon_i} \tag{1}$$

$$t_{ij}^{s} = \frac{2k_{iz}}{k_{iz} + k_{jz}}, \qquad t_{ij}^{p} = \frac{2k_{iz}n_i n_j}{k_{iz}\epsilon_j + k_{jz}\epsilon_i} \tag{2}$$

The diagonal Fresnel transformation tensor is an efficient way to express Fresnel reflection coefficients in compact form as

$$\overleftrightarrow{r}_{ij} = r_{ij}^{xx}\hat{x}\hat{x} + r_{ij}^{yy}\hat{y}\hat{y} + r_{ij}^{zz}\hat{z}\hat{z} = -r_{ij}^{p}\hat{x}\hat{x} + r_{ij}^{s}\hat{y}\hat{y} + r_{ij}^{p}\hat{z}\hat{z} \tag{3}$$

Here, we denote the incident fundamental field at frequency $\omega$ propagating in free space (medium 1) at an incident angle $\theta$

$$\boldsymbol{E}_{in}(\omega, \boldsymbol{r}) = E_{in} e^{ik_1 . r}\hat{e}_{in} \tag{4}$$

where $\boldsymbol{r}(x, y, z)$ represent the position vector in space and $\hat{e}_{in} = \boldsymbol{E}_{in}/|\boldsymbol{E}_{in}|$ is the normalized electric field vector with $\hat{e}_{in}. k = 0$.

After considering multiple reflections of the fundamental field within the thin film, local fundamental field at the surface at $\boldsymbol{r} = (0,0,0)$ results in

$$\boldsymbol{e_s}(\omega) = \frac{\left(\overleftrightarrow{1} + \overleftrightarrow{r}_{12}\right)\left(\overleftrightarrow{1} + \overleftrightarrow{r}_{23} e^{2ik_{2z}d}\right)}{\overleftrightarrow{1} - \overleftrightarrow{r}_{21}\,\overleftrightarrow{r}_{23} e^{2ik_{2z}d}}\hat{e}_{in} \tag{5}$$

Similarly, the local fundamental field at the interface at $\boldsymbol{r} = (0,0,d)$ will be represented as

$$\boldsymbol{e_I}(\omega) = \frac{\left(\overleftrightarrow{1} + \overleftrightarrow{r}_{12}\right)\left(\overleftrightarrow{1} + \overleftrightarrow{r}_{23}\right) e^{ik_{2z}d}}{\overleftrightarrow{1} - \overleftrightarrow{r}_{21}\,\overleftrightarrow{r}_{23} e^{2ik_{2z}d}}\hat{e}_{in} \tag{6}$$

Now, diagonal tensors $\overleftrightarrow{f_s}(\omega)$ and $\overleftrightarrow{f_I}(\omega)$ are defined for representing local normalized fundamental fields described in eqn. (5) and (6) in a compact form. Thus,

$$\boldsymbol{e_L}(\omega) = \overleftrightarrow{f_L}\hat{e}_{in} \tag{7}$$

where $L \in \{S, I\}$ and $\overleftrightarrow{f_L} = f_L^{xx}\hat{x}\hat{x} + f_L^{yy}\hat{y}\hat{y} + f_L^{zz}\hat{z}\hat{z}$ with components

$$\overleftrightarrow{f_s}^{\xi\xi} = \frac{\left(1 + \overleftrightarrow{r}_{12}^{\xi\xi}\right)\left(1 + \overleftrightarrow{r}_{23}^{\xi\xi} e^{2ik_{2z}d}\right)}{1 - \overleftrightarrow{r}_{21}^{\xi\xi}\,\overleftrightarrow{r}_{23}^{\xi\xi} e^{2ik_{2z}d}} \tag{8}$$

$$\overleftrightarrow{f_I}^{\xi\xi} = \frac{\left(1 + \overleftrightarrow{r}_{12}^{\xi\xi}\right)\left(1 + \overleftrightarrow{r}_{23}^{\xi\xi}\right) e^{ik_{2z}d}}{1 - \overleftrightarrow{r}_{21}^{\xi\xi}\,\overleftrightarrow{r}_{23}^{\xi\xi} e^{2ik_{2z}d}} \tag{9}$$

where $\xi \in \{x, y, z\}$ and $r_{ij}^{\xi\xi}$ represent the components of the Fresnel transformation tensor for reflection as described in eqn. (3).

SHG radiation generated at $r = (0,0,0^-)$ also experiences multiple reflections similar to the fundamental field. Therefore, local field relations described in eqn. (8) and (9) are still valid for SHG fields at the surface and interface. Upper case notation is chosen for tensors describing multiple reflections and local fields of SHG making it distinct from fundamental field.

Analogously, propagation of SHG electric field at surface or interface with polarization vector $\hat{e}_{out}$ can be evaluated as

$$\boldsymbol{e_L}(2\omega) = \hat{e}_{out}.\overleftrightarrow{F}_L \tag{10}$$

Now, SHG electric field radiated from the polarization sheet is projected onto the polarization vector and can be expressed as

$$\hat{e}_{out}.\boldsymbol{E_L}(2\omega) = \frac{i\omega}{c\cos(\theta)}\boldsymbol{e_L}(2\omega).\overleftrightarrow{\chi}_L^{(2)}(2\omega){:}\,\boldsymbol{e_L}(\omega)\boldsymbol{e_L}(\omega)E_{in}^2(\omega) \tag{11}$$

This leads to the total SHG intensity by following expression

$$I(2\omega) = \frac{2\omega^2}{\epsilon_0\, c^3\cos^2\theta}\left|\sum_L \boldsymbol{e_L}(2\omega).\overleftrightarrow{\chi}_L^{(2)}(2\omega){:}\,\boldsymbol{e_L}(\omega)\boldsymbol{e_L}(\omega)\right|^2 I_{in}^2(\omega) \tag{12}$$

Eqn. (12) is employed in numerical calculation of SHG at the surface and interface of thin film. Moreover, summation of both contributions represents the total surface-induced SHG.

### 2.2 Transfer matrix method

In thin-film optics, TMM is a widely adopted framework for the investigation of light propagation in 1D multilayer structures, including one-dimensional (1D) photonic crystals, Bragg reflectors [35]. Beyond its traditional applications in linear optics, TMM has utilized extensively to examine the nonlinear phenomena in layered media [25-32].

This approach commonly relies on the electromagnetic boundary conditions in homogeneous, isotropic and linear medium as described in Ref. [35]. But these boundary conditions are inadequate when a surface is subjected to a high-intensity optical field [32,36]. In such cases, nonlinear source polarization is generated at the surface, which emits electromagnetic radiation giving rise to jump in the tangential component of the electric and magnetic field. Therefore, boundary conditions are modified and takes generalized form that can be expressed as

$$\Delta E_{||} = -\frac{1}{\epsilon_0 \epsilon_2} \vec{\nabla}_{||} P_{S,\perp}^{NL} \tag{13}$$

$$\Delta H_{||} = 2i\omega \vec{P}_S^{NL} \times \vec{r}_\perp \tag{14}$$

where $\epsilon_0$, and $\epsilon_2$ denotes the dielectric permittivity of free space and the thin film, respectively, while $\omega$ is the fundamental frequency. $\vec{P}_S^{NL}$ denotes the surface nonlinear polarization and $\perp, ||$ represent the direction perpendicular and parallel to the interface. $\vec{\nabla}_{||}$ represents the gradient in the plane parallel to the interface i.e. tangential plane and $\Delta$ stands for the discontinuity at the respective plane. These, boundary conditions introduce an additional source matrix in usual TMM [32].

For a particular case of linear polarized light where electric field is oriented along $x$ direction and magnetic field is in $y$ direction, boundary conditions allow the electric fields in substrate (region 3) and superstrate (region 1) to be connected through the following equality

$$\begin{bmatrix} E_x \\ H_y \end{bmatrix}_3 = M \begin{bmatrix} E_x \\ H_y \end{bmatrix}_1 + S \tag{15}$$

where $M$ represents the transfer matrix and $S$ denotes the nonlinear source polarization matrix or source matrix. The source matrix $S = S_{12} + M * S_{01}$ where $S_{12}$ is the source matrix at the film-substrate interface and $S_{01}$ denotes the source matrix at air-film interface. Importantly, the formulation of the transfer matrix ($M$) significantly depends upon the polarization of incident light, whether it is TE or TM polarized. The source matrix solely depends upon the boundary conditions which are inherently connected to the induced polarization vector. Thus, symmetry of the medium can affect induced polarization, which eventually influences the induced SHG. The source matrix can be expressed as

$$S = \begin{bmatrix} -\Delta E_{||} \\ -\Delta H_{||} \end{bmatrix} \tag{16}$$

### 2.2.1 For TE or s-polarized wave

First, we assume TE-polarized SHG light emitted from the thin film. According to our geometry, a TE-polarized light contains an electric field in $y$ direction, while magnetic field lies in the plane of incidence, i.e. $x - z$ plane. Subsequently, eqn. (15) can be expressed as follows

$$\begin{bmatrix} E_{TE,y} \\ H_{TE,x} \end{bmatrix}_3 = M_{SH}^{TE} \begin{bmatrix} E_{TE,y} \\ H_{TE,x} \end{bmatrix}_1 + S^{TE} \tag{17}$$

where $M_{SH}^{TE}$ represents the transfer matrix at second harmonic frequency and $S^{TE}$ is known as nonlinear source polarization matrix whose elements depends upon the generalized boundary conditions described in eqn. (1) and (2). In this case, transfer matrix for a thin layer of thickness $d$ can be written as

$$M_{SH}^{TE} = \begin{bmatrix} \cos\left(k_{2,z}^{SH} d\right) & iZ' \sin\left(k_{2,z}^{SH} d\right) \\ \frac{i}{Z'} \sin\left(k_{2,z}^{SH} d\right) & \cos\left(k_{2,z}^{SH} d\right) \end{bmatrix} \tag{18}$$

where $Z' = Z\frac{k}{k_z}$ and parameter $k, k_z$ denotes the propagation vector and projection of the wave vector in $z$ direction respectively. Parameter $Z = \sqrt{\frac{\mu_0}{\epsilon_0}}$, where $\mu_0$ and $\epsilon_0$ are the permeability and permittivity of the free space. The superscript SH denotes the parameters at second harmonic frequency.

Incident plane wave generates two plane waves after encountering the thin film propagating towards superstrate and substrate region and their amplitudes can be denoted as $A^+$ and $A^-$ respectively. Thus, electric and magnetic field relation in a plane wave provides a relation

$$\begin{bmatrix} E_{TE,y} \\ H_{TE,x} \end{bmatrix}_3 = \begin{bmatrix} A^- \\ A^-/Z_3' \end{bmatrix} \tag{19}$$

$$\begin{bmatrix} E_{TE,y} \\ H_{TE,x} \end{bmatrix}_0 = \begin{bmatrix} A^+ \\ -A^+/Z_1' \end{bmatrix} \tag{20}$$

Now, substitute eqn. (19) and (20) in eqn. (17) and obtain following system of equations with unknowns $A^+$ and $A^-$

$$\begin{bmatrix} A^- \\ A^-/Z_3' \end{bmatrix} = M^{TE} \begin{bmatrix} A^+ \\ -A^+/Z_1' \end{bmatrix} + S^{TE} \tag{21}$$

By solving coupled equations obtained from eqn. (21), yields the final expression for the amplitude in the superstrate region

$$A^+ = \frac{S_2^{TE} Z_1' Z_3' - S_1^{TE} Z_1'}{M_{11} Z_1' + M_{22} Z_3' - M_{12} - M_{21} Z_1' Z_3'} \tag{22}$$

### 2.2.2 For TM polarized light or p-polarized wave

This section starts with the assumption of a TM-polarized second harmonic wave from thin film being detected. In a TM-polarized emitted second harmonic wave, the electric field lies in $xy$-plane while magnetic field is oriented along $y$ direction. Therefore, eqn. (15) takes the following form

$$\begin{bmatrix} E_{TM,x} \\ H_{TM,y} \end{bmatrix} = M_{SH}^{TM} \begin{bmatrix} E_{TM,x} \\ H_{TM,y} \end{bmatrix} + S^{TM} \tag{23}$$

where $M_{SH}^{TM}$ represents the transfer matrix at second harmonic frequency and $S^{TM}$ is the source polarization matrix. In the case of TM-polarized light, transfer matrix is expressed as follows

$$M_{SH}^{TM} = \begin{bmatrix} \cos\left(k_{2,z}^{SH} d\right) & -iZ'' \sin\left(k_{2,z}^{SH} d\right) \\ -\frac{i}{Z''} \sin\left(k_{2,z}^{SH} d\right) & \cos\left(k_{2,z}^{SH} d\right) \end{bmatrix} \tag{24}$$

where $Z'' = Z \frac{k_z}{k}$ and all the parameters have the same meaning as defined in TE case. Here, the $x$ component of the electric field in the superstrate and substrate region is denoted by $A^+$ and $A^-$ respectively. Subsequently, relation between electric and magnetic field in a plane wave allow an equality relation which is given by

$$\begin{bmatrix} E_{TM,x} \\ H_{TM,y} \end{bmatrix}_3 = \begin{bmatrix} A^- \\ -A^-/Z_3'' \end{bmatrix} \tag{25}$$

$$\begin{bmatrix} E_{TM,x} \\ H_{TM,y} \end{bmatrix}_1 = \begin{bmatrix} A^+ \\ A^+/Z_3'' \end{bmatrix} \tag{26}$$

Now, we substitute eqn. (25) and (26) in eqn. (23) and a system of equations is obtained with unknowns $A^+$ and $A^-$ that can be described as

$$\begin{bmatrix} A^- \\ -A^-/Z_3'' \end{bmatrix} = M^{TM} \begin{bmatrix} A^+ \\ A^+/Z_3'' \end{bmatrix} + S^{TM} \tag{27}$$

After solving eqn. (27) expression for the amplitude of SHG field reflected in the superstrate region is obtained that is

$$A^+ = \frac{-S_2^{TM} Z_0'' Z_3'' - S_1^{TM} Z_0''}{M_{11} Z_0'' + M_{22} Z_3'' + M_{12} + M_{21} Z_0'' Z_3''} \tag{28}$$

### 2.2.3 Calculation of the source term

To evaluate SHG field amplitude using eqn. (22) and eqn. (28), source matrices $S^{TE}$and $S^{TM}$ has to be calculated based on the boundary conditions described in eqn. (13), (14) and (16).

For this purpose, we consider thin polarization sheet at each interface of the thin film, where surface nonlinear polarization is expressed through the initial parameters of incident fundamental wave. In this process, we neglect the reverse action of SHG field on incident fundamental wave, which is known as undepleted pump approximation.

The surface nonlinear polarization at the homogenous, isotropic surface can be expressed as

$$P_s^{NL} = \epsilon_0 \chi^{(2)}_{\perp\perp\perp} E_\perp(\omega,\vec{r}) E_\perp(\omega,\vec{r}).\vec{n} + \epsilon_0 \chi^{(2)}_{\perp || ||} E_{||}(\omega,\vec{r}) E_{||}(\omega,\vec{r}).\vec{n} + \epsilon_0 \chi^{(2)}_{||\perp||} E_\perp(\omega,\vec{r}) E_{||}(\omega,\vec{r}).\vec{\tau} \quad (29)$$

where $\vec{n}$ is the outward normal unit vector and $\vec{\tau}$ is the tangent unit vector at point $r$. $E_\perp$ and $E_{||}$ are the normal and tangential components of the fundamental field. Here, nonlinear polarization is demonstrated for only isotropic surfaces, while other terms may arise for non-centrosymmetric materials, and susceptibility components can be determined by the crystal symmetry. Moreover, eqn. (29) involves both $E_\perp$ (normal) and $E_{||}$ (tangential) components of the electric field of incident fundamental wave; therefore, these components have to be computed at each surface of the thin film.

We start with a TE-polarized incident plane wave at the fundamental frequency and the amplitude of the reflection coefficient from a thin film is denoted as $R_{TE}$. The electric field amplitude of the incident fundamental wave is $E_I$. Then, our geometry suggests that $\vec{k}_{||}^{FW}$ and $\vec{k}_{||}^{SHG}$ are directed along $x$ axis and the electric field component of incident TE-polarized fundamental wave should be along the $y$ axis, while the magnetic field lies in the $xz$ plane. Then, TMM described in eqn. (18) is rewritten for the fundamental frequency and the fundamental field right after the surface can be evaluated using the following expression

$$\begin{bmatrix} E_{TE,y} \\ H_{TE,x} \end{bmatrix} = M_{FW}^{TE} \begin{bmatrix} E_I(1+R_{TE}) \\ E_I \dfrac{1-R_{TE}}{Z'_{1,FW}} \end{bmatrix} \quad (30)$$

Here, $M_{FW}^{TE}$ is the transfer matrix of the film at fundamental frequency and subscript $FW$ denotes the parameter calculated at fundamental frequency.

A similar analytical procedure can be performed for the TM-polarized fundamental wave where magnetic field is oriented in $y$ direction while the electric field lies in the $xz$ plane. So, tangential component is oriented in $x$ direction. In this case, the amplitude of the reflection coefficient is defined as $R_{TM}$ while the amplitude of tangential component of incident fundamental wave is $E_{I,x}$. Thus, electric field just after the surface can be written as

$$\begin{bmatrix} E_{TM,x} \\ H_{TM,y} \end{bmatrix} = M_{FW}^{TM} \begin{bmatrix} E_{I,x}(1+R_{TM}) \\ -E_{I,x} \dfrac{1-R_{TM}}{Z_{1,FW}} \end{bmatrix} \quad (31)$$

Here, $M_{FW}^{TM}$ is the transfer matrix of the film for TM polarized light at fundamental frequency.

Now, the normal component of the electric field can be evaluated by using the tangential component of the magnetic field and these quantities are connected through the following relation

$$E_{TM,z} = -\frac{k_{m+1,x}^{FW}}{k_{m+1}^{FW}} Z_{m+1,FW} H_{TM,y} \quad (32)$$

where $m = 0$ and 1 describe the electric field just after the surface 1 and interface 2.

To evaluate the electric field using eqn. (30) and (31), the amplitude of reflection coefficient has to be computed first. So, we follow the process given in ref. [32,35] and obtain the expression for $R_{TE}$ which is given as follows

$$R_{TE} = \frac{M_{22}Z'_{3,FW} - M_{11}Z'_{0,FW} + M_{21}Z'_{0,FW}Z'_{3,FW} - M_{12}}{M_{22}Z'_{3,FW} + M_{11}Z'_{0,FW} - M_{21}Z'_{0,FW}Z'_{3,FW} - M_{12}} \tag{33}$$

Similar procedure is followed for the TM polarized incident fundamental light and amplitude of reflection coefficient is expressed as

$$R_{TM} = \frac{M_{22}Z''_{3,FW} - M_{11}Z''_{0,FW} - M_{21}Z''_{0,FW}Z''_{3,FW} + M_{12}}{M_{22}Z''_{3,FW} + M_{11}Z''_{0,FW} + M_{21}Z''_{0,FW}Z''_{3,FW} + M_{12}} \tag{34}$$

### 2.3 $\chi^{(2)}$ rescaling

$\chi^{(2)}$ rescaling plays an important role in achieving identical SHG intensity using both frameworks and establishing equivalence between them. If both methods calculate the nonlinear source differently, the nonlinear susceptibilities in one method must be converted with respect to the second method.

The exact position of the electric field generating the SHG can be placed in three different positions: inside the film, outside the film, or in a vacuum gap. The tangential component of the electric field is continuous across the boundary, while the normal component encounters discontinuity at the boundary. Therefore, induced nonlinear polarization involving the normal component of the fundamental electric field must use rescaled nonlinear susceptibility components if the fundamental fields are defined differently. In addition, nonlinear susceptibilities should also be rescaled if the position of the nonlinear source of polarization sheet is different in both methods. Here, we describe the process of rescaling of nonlinear susceptibility components for the methods described in previous sections.

In the multiple reflection model, electric fields are calculated in the vacuum gap at the surface and interface, while TMM employs the electric field inside the thin film. Thus, the normal component of the fundamental electric field is connected through the relation $E^{TMM} = \frac{E^{Geilis}}{\epsilon_2}$, while the tangential component remains continuous.

Moreover, the denominator of eqn. (13) includes $\epsilon_{2,SH}$ demonstrating the position of the source polarization sheet inside the thin film at frequency $2\omega$. On the other contrary, the multiple reflection method assumes a nonlinear source polarization sheet in the vacuum gap at the surface and interface. Therefore, susceptibility needs to be rescaled, taking the following form: $\chi_{zzz}^{TMM} = \frac{\chi_{zzz}^{Geilis}}{\epsilon_{2,SH}\,\epsilon_2^2}$, $\chi_{zxx}^{TMM} = \frac{\chi_{zxx}^{Geilis}}{\epsilon_{2,SH}}$ and $\chi_{xzx}^{TMM} = \frac{\chi_{xzx}^{Geilis}}{\epsilon_2}$.

It is worth noting that the scaling above is valid only for the particular presented case. If the position of the nonlinear polarization sheet is changed, the denominator of eqn. (13) will be changed and subsequently $\epsilon_{2,SH}$ will be replaced with that particular permittivity value. Analogously, the scaling of the susceptibility tensor will change.

## 3. Results and Discussion

In this section, we present numerical simulations based on methods described in Section 2.1 and 2.2 to compare their respective results. The system under consideration consists of a single layer of thickness $1200$ nm deposited on a semi-infinite substrate. A fundamental wave with wavelength $1030$ nm is launched from air into the thin film at an incident angle $\theta$ generating second harmonic wave at wavelength $515$ nm. The refractive index of the medium at the fundamental wavelength given as $n_1 = 1$, $n_2 = 2.5$ and $n_3 = 1.5$ while at second harmonic $n_1 = 1$, $n_2 = 2.7$ and $n_3 = 1.5$. Moreover, the components of surface and interface susceptibilities are $\chi_{zzz}^{surf} = 1$, $\chi_{zxx}^{surf} = 0.2$, $\chi_{xzx}^{surf} = 0.4$ and $\chi_{zzz}^{int} = 1$, $\chi_{zxx}^{int} = 0.5$, $\chi_{xzx}^{int} = 0.8$ respectively which are used in our TMM calculations and the rescaled values are utilized in the multiple reflection model.

All the calculated amplitudes and intensities are in dimensionless form. Here, all the physical parameters are dimensionless i.e. normalized.

The choice of sign convention must remain consistent across the analytical expressions derived from both methods. This consistency is important when TM polarized fundamental wave is considered as the orientation and sign of the normal component of electric field directly affect the induced nonlinear polarization, as we show later in the text.

### 3.1 Numerical discrepancies in the method comparison

In this section, we discuss the origin of potential discrepancies in the numerical implementation of both theoretical frameworks, which can lead to the ambiguous total SHG intensity of the thin film, despite using the same input parameters. While both methods can predict identical results under certain conditions that will be discussed in the following subsections, here we focus on the contradictory results.

#### 3.1.1 Position of nonlinear polarization sheet

In a centrosymmetric thin film, inversion symmetry is broken only at the surface and buried interface of the thin film. Therefore, the incident fundamental wave induces nonlinear polarization only at the surfaces, which is considered as an infinitesimally thin sheet. The position of this polarization sheet plays a vital role in the emitted SHG radiation. Therefore, strict consistency is needed in the implementation of the induced nonlinear polarization sheet.

In our single-layer system, the polarization sheet can be assumed at the following three positions, 1. Inside the thin film 2. Vacuum gap at the surface and buried interface of the film 3. Outside of the thin film (superstrate and substrate). Among these possibilities, the position can be chosen arbitrarily, but it needs to be followed in both theoretical models. In addition, the nonlinear surface susceptibilities values are rescaled in the multiple-reflection model and the source polarization is modified in TMM.

We demonstrate the need for a consistent approach in Fig. 2, where we show the total SHG intensity from a layer when both the fundamental and SHG waves are TM-polarized. When the polarization sheet is assumed to lie inside the thin-film medium for both methods, and the same nonlinear susceptibilities are used in both methods, i.e. $\chi_{zzz}^{TMM} = \chi_{zzz}^{Geilis}, \chi_{zxx}^{TMM} = \chi_{zxx}^{Geilis}, \chi_{xzx}^{TMM} = \chi_{xzx}^{Geilis}$, the total SHG intensity calculated from both methods disagree as shown in Fig. 2 (a).

In the second example in Fig. 2(b), we use rescaled nonlinear susceptibilities as described in the previous section, but the location of the polarization sheet is inconsistent: it is assumed to lie inside thin film in the multiple-reflection model while TMM considers it in vacuum gap. Also in this case, the total SHG disagrees.

Finally, in Fig. 2 (c), we placed the nonlinear polarization sheet just before both the surface and interface in TMM while keeping it inside thin film in the multiple-reflection method. Also in this configuration, the resultant total SHG angle dependence obtained from both methods is not in agreement.

On the other hand, a slightly different scenario can be observed when a TE-polarized incident light is taken into consideration. The total SHG intensity pattern remains unchanged if one method assumes a polarization sheet either inside thin film or in vacuum gap with respect to other method. Such behavior arises because for TE-polarized light, only one susceptibility component contributes to the SHG generation. Therefore, the shape of the SHG angle dependence remains intact from interference originating from the contribution of other susceptibility components.

Fig. 3 demonstrates that even for TE-polarized light, total SHG intensity can differ if one method selects the polarization sheet inside thin film or in vacuum, while second method prefer polarization sheet just before the surface and interface of the film.

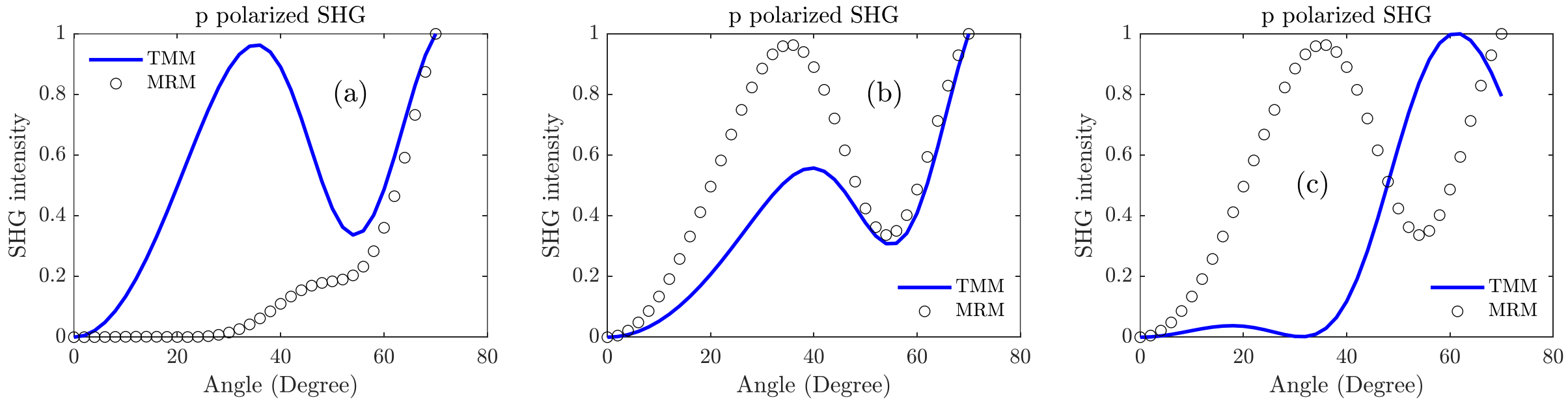

Figure 2. Discrepancies in total SHG intensity when (a) susceptibilities are not rescaled in multiple reflection model and polarization sheet is inside thin film (b) polarization sheet are in vacuum gap in TMM and within film in multiple reflection model (c) polarization sheet just right before the surface and interface (see schematic) in TMM and within film in multiple reflection model

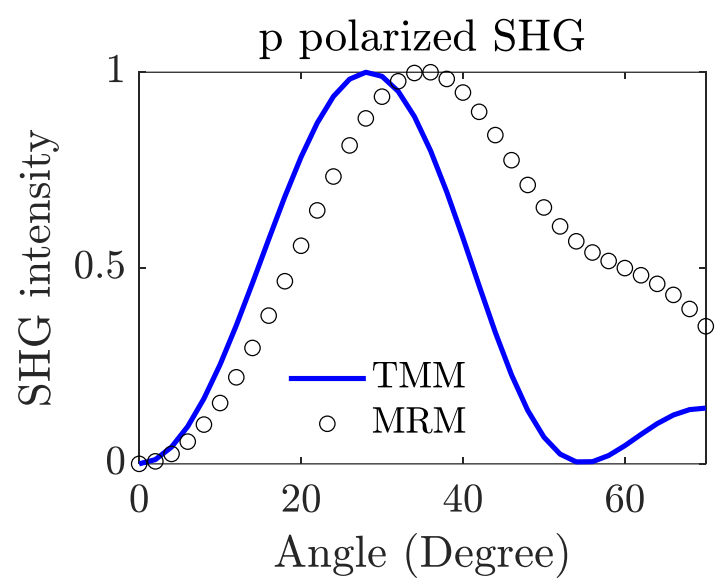

Figure 3. Total SHG from a thin film when a TE polarized light is launched and polarization sheet is considered inside thin film in one method while before surface and interface in other method.

### 3.1.2 Reference plane of SHG radiation

In TMM, the fundamental wave induces nonlinear polarization at both the surface and the interface, which act as a source for the second harmonic wave. The source field generated at the air-film surface ($S_{01}$) propagates to the film-substrate interface ($S_{12}$) then total addition of these sources is utilized in the calculation of the generated second harmonic wave. Thus, the total source field is represented by $S = S_{12} + M_{SH} * S_{01}$ where $M_{SH}$ is included for incorporating the propagation of $S_{01}$ from surface to interface – see Section 2.2. If we impose the reverse assumption and calculate total source field at the air-film surface by using $S = M_{SH} * S_{12} + S_{01}$ then the resultant total SHG will disagree between the two methods as shown in Fig. 4.

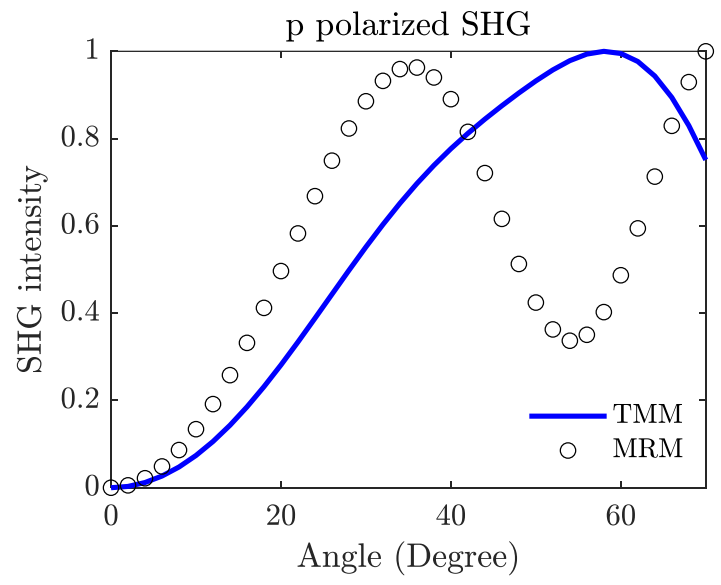

Fig. 4. Total SHG intensity from thin film when TM polarized light is launched and reference plane is shifted to air-film surface by propagating interface emitted field to surface.

### 3.1.3 Projection of input and output light

In the multiple-reflection model, the local electric field of the fundamental wave is calculated at the surface and interface and, subsequently, the x/y/z-projections of the p/s-polarized field are used to calculate the induced nonlinear polarization. On the other hand, the output SHG field in x/y/z components is recalculated into the s/p-polarization by the projection of the emitted field at second harmonic frequency. In these projections, only projections consistent of fundamental and second harmonic waves, that are consistent in their signs, allow results that agree well with TMM particularly in TM polarized incident fundamental wave. Fig. 5 shows that inappropriate projection governs the distinct total SHG intensity in the multiple-reflection model, leading to a clear disagreement with TMM results when the incident fundamental wave is TM polarized.

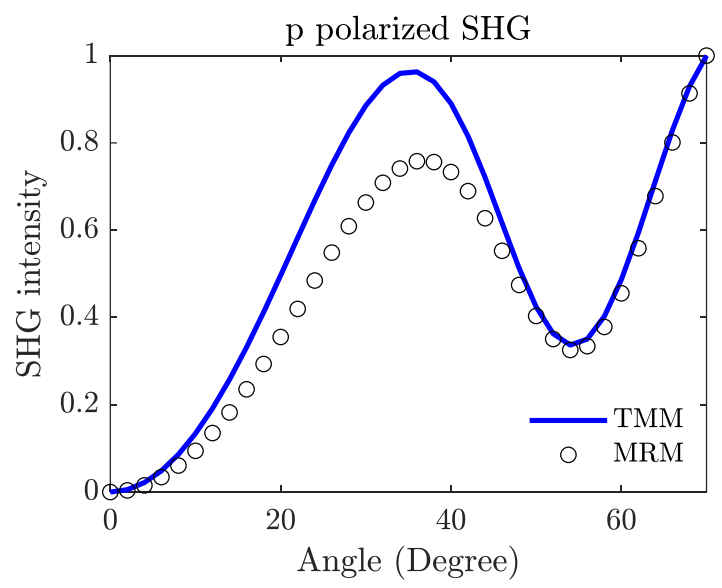

Figure 5. Total SHG intensity when TM-polarized light is launched onto a thin film and incorrect projections are taken

### 3.2. Consistent theory implementation

The results above disclosed the reasons of potential disagreement in the total SHG obtained from both methods that could raise a concern on the theoretical consistency of these frameworks. By considering these facts, we will demonstrate in this subsection a consistent numerical implementation of both methods and compare them systematically. We ensure that all these discrepancies in the total SHG can be removed by adopting a consistent numerical implementation.

First, we start with TE polarized fundamental wave and subsequently the TM polarized will be discussed.

As it became clear from the results, it is important to keep the same nonlinear polarization sheet position (within thin film) and rescale the nonlinear susceptibilities for this case: $\chi_{zzz}^{TMM} = \frac{\chi_{zzz}^{Geilis}}{\epsilon_{2,SH}\,\epsilon_2^2}$, $\chi_{zxx}^{TMM} = \frac{\chi_{zxx}^{Geilis}}{\epsilon_{2,SH}}$, $\chi_{xzx}^{TMM} = \frac{\chi_{xzx}^{Geilis}}{\epsilon_{2,}}$ where $\epsilon_2$ and $\epsilon_{2,SH}$ are the dielectric constants of the thin film at the fundamental wavelength and the second-harmonic wavelength, respectively. Moreover, $\epsilon_{2,SH}$ is used in the denominator of eqn. (13) to make the source matrix consistent.

#### 3.2.1 TE-polarized fundamental wave

An incident TE-polarized fundamental wave exhibits $H_x, E_y, H_z$ as nonzero field components where the electric field $E_y$ is oriented in $y$ direction. Thus, being a tangential component of the fundamental electric field, $E_y$ must be continuous across the boundary. Since the fundamental field at the surface is the first step in the calculation of surface SHG, it is essential to be compared. Fig. 6 shows the comparison of $E_y$ at the surface and interface of the thin film.

Here, $E_y$ obtained from both methods is identical. As shown in Fig. 6 (a) and (b), real and imaginary parts of the electric field calculated at the air-film surface (i.e. 01 surface) is same. Moreover, the phase of this electric field is also identical in both methods, as depicted in Fig. 6 (c). Similarly, both methods provide identical electric field at film-substrate interface (i.e. 12 surface) as demonstrated in Fig. 6 (d) and (e) where the real and imaginary parts of the electric field are found to be the same. Hence, the phase accumulated by the propagated field is also identical, as shown in Fig. 6 (f).

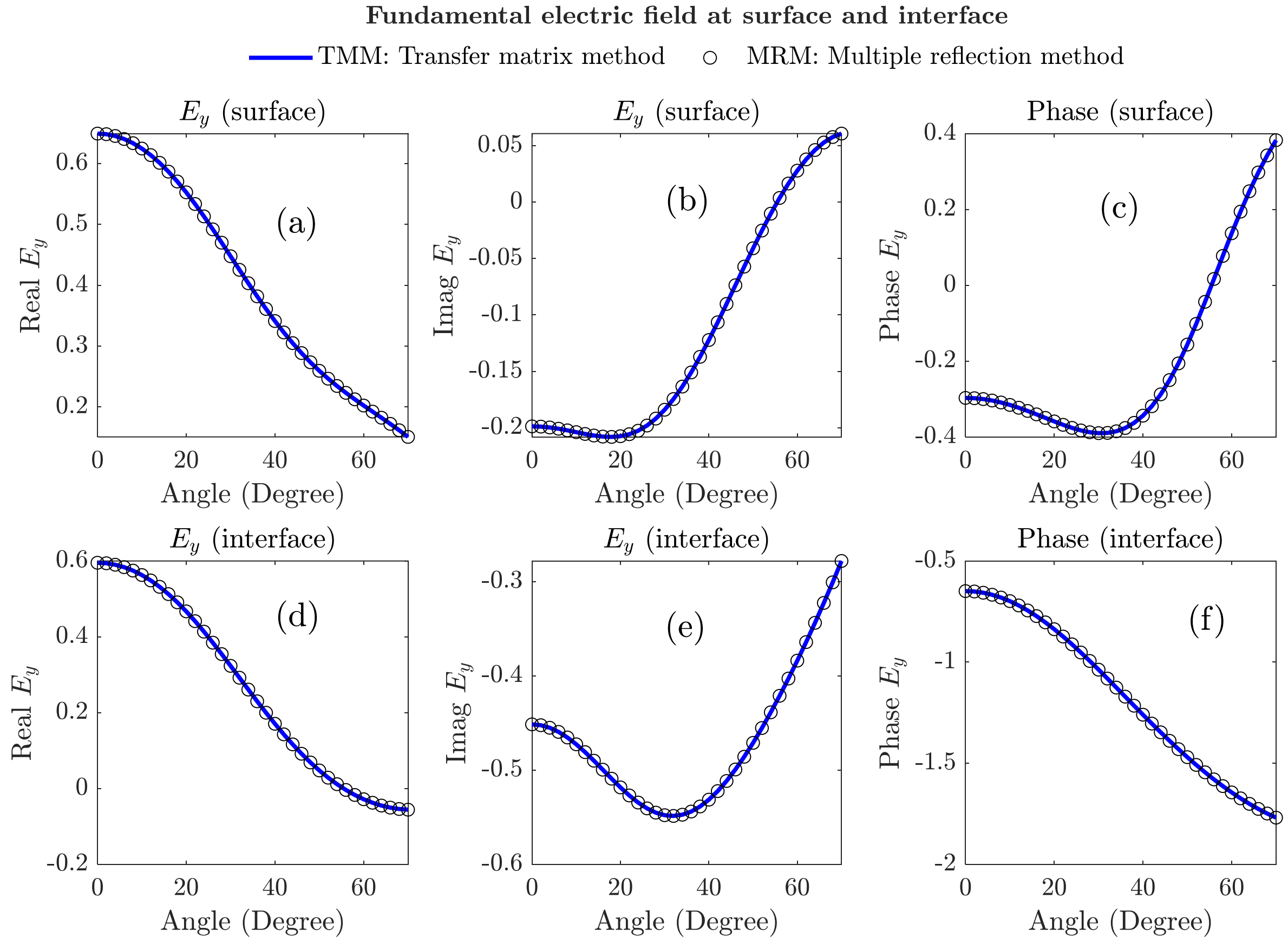

Figure 6. (a), (d) Real (b), (e) imaginary part of the electric field and (c), (f) accumulated phase with incident angle from both methods at (upper row) air-thin film (surface) boundary (lower row) thin film substrate boundary (interface)

The electric field of the fundamental wave is employed in the calculation of induced nonlinear polarization. In our system under consideration, the layer surface is assumed to be homogeneous, isotropic, and $C_\infty$ symmetric therefore only $z$ component of the nonlinear polarization is contributing to the radiated SHG field. Therefore, $P_z$ remains the only nonzero component, while $P_x$ and $P_y$ becomes zero in this case.

Fig. 7 illustrates the induced nonlinear polarization profiles and their accumulated phases at the surface and interface of the thin film. Specifically, the real and imaginary part of the polarization components $P_z$ are plotted in Fig. 7 (a) and (b), respectively, while the angular dependence of the phase of $P_z$ is demonstrated in Fig. 7 (c). These quantities confirm the strict equivalence of induced nonlinear polarization calculated at the surface using both methods. Analogous calculations performed for the interface led to identical results, which are demonstrated in Fig. 7 (d)-(f).

Induced nonlinear polarization dictates the radiated SHG fields. Therefore, the angular dependence of the phase must be identical across both methods.

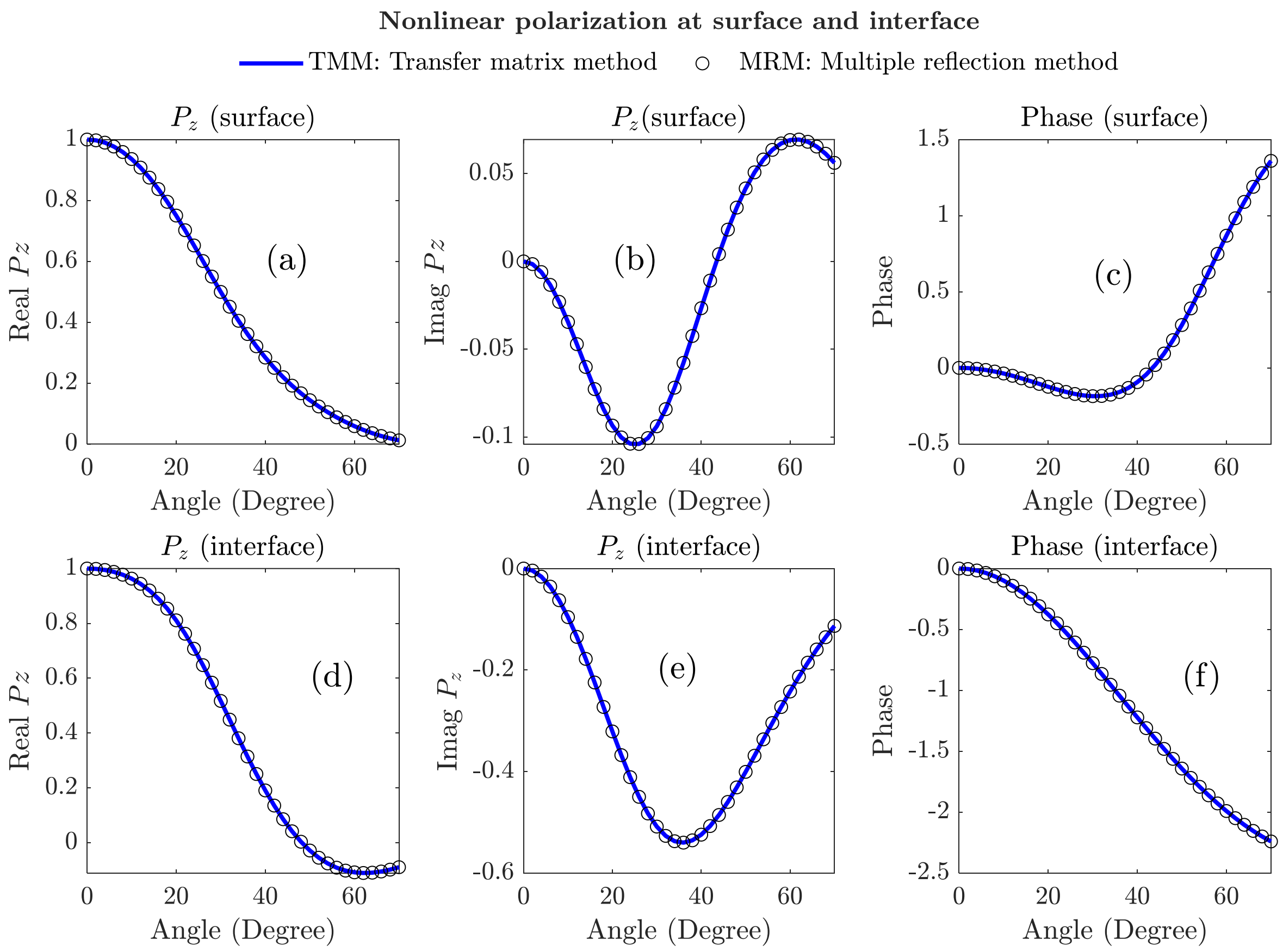

Figure 7. (a), (d) Real (b), (e) imaginary part of induced nonlinear polarization; and (c), (f) accumulated phase with incident angle from both methods at (upper row) air-thin film (surface) boundary (lower row) thin film substrate boundary (interface)

Identical fundamental field and induced nonlinear polarization pave the way to calculate SHG field using these quantities. The induced nonlinear polarization at the surface and interface acts as a source field radiating emission at the second-harmonic wavelength. Fig. 8 (a) and 8 (b) show the real and imaginary parts of the total SHG electric field, which results from the interference between SHG electric fields generated at the surface and interface. Interference of SHG electric fields is strongly influenced by the phase accumulated by the constituent fields. Hence, the SHG electric fields generated at the surface and interface must exhibit identical phases in both methods; otherwise resulting total SHG electric field would differ upon interference. As shown in Fig. 8 (c), the total phase of total SHG electric field radiated from film is completely consistent in both methods. The total SHG field subsequently used in the calculation of total SHG intensity as shown in fig. 9 (a) where both methods are shown to predict identical total SHG intensity from our thin film system. Moreover, fig. 9 (b) demonstrate that reflectivity of the fundamental wave remains invariant irrespective of the methods is being used.

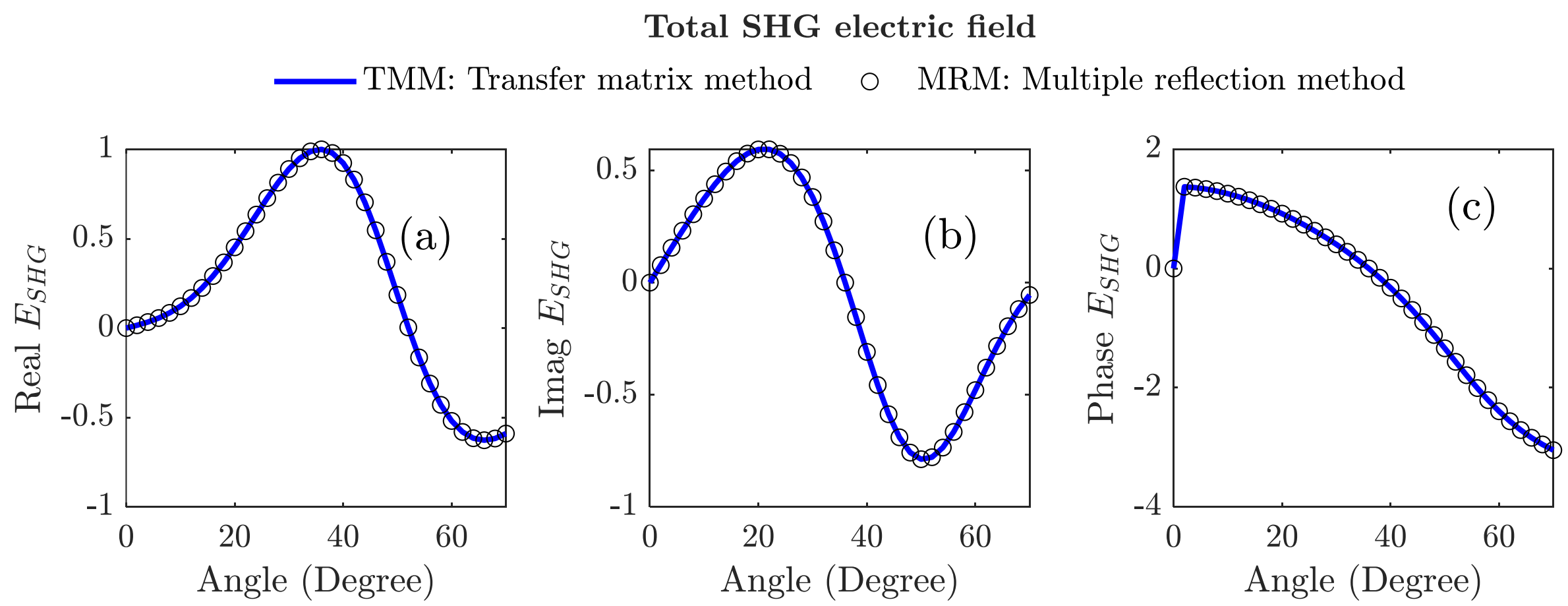

Figure 8 (a) Real (b) imaginary part of the second harmonic electric field and (c) accumulated phase by the induced SHG with the variation of incident angle

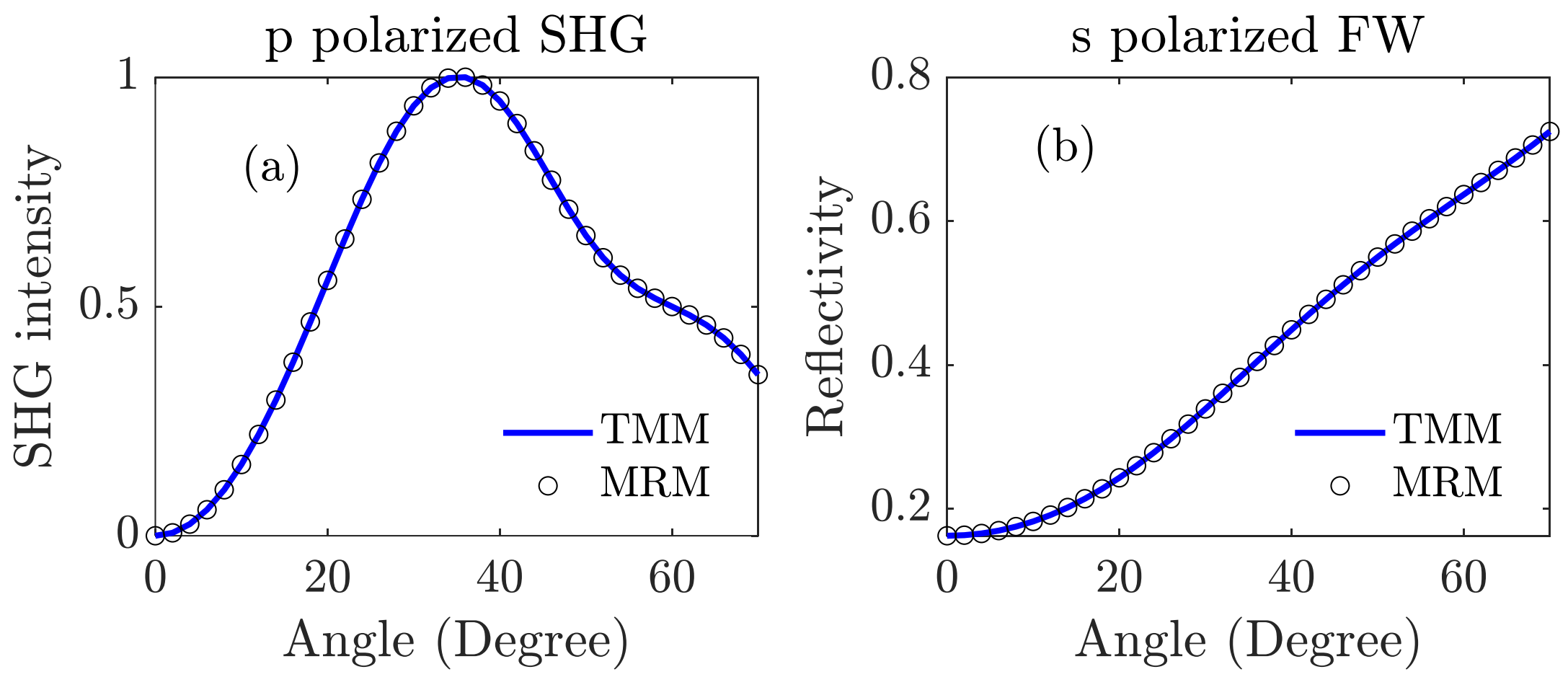

Figure 9 (a) Total intensity of SHG (surface + interface) and (b) reflectivity of the incident FW wave

### 3.3 TM polarized fundamental wave

Now, we consider the incident fundamental wave to be TM-polarized; then it must have $E_x, H_y, E_z$ nonzero field components according to our geometry as shown in Fig. 1. In this case, $E_x,\ E_z$ are the tangential and normal components of the fundamental field, respectively. Therefore, the induced nonlinear polarization will be modified according to the eqn. (29), which eventually modifies the SHG electric field.

First, the tangential component of the electric field ($E_x$) is calculated and found to be identical across both methods. The exact equivalence between real and imaginary parts of the $E_x$ is demonstrated in Fig. 10 (a) and (b), which underpin the consistent angular dependence of its phase depicted in Fig. 10 (c). Similarly, the normal component ($E_z$) of the fundamental electric field at the surface and its corresponding phase evolution are found identical in both methods as shown in Fig. 11 (a)-(c). This equivalence ensures that both approaches can provides identical nonlinear polarization when the incident wave is TM polarized.

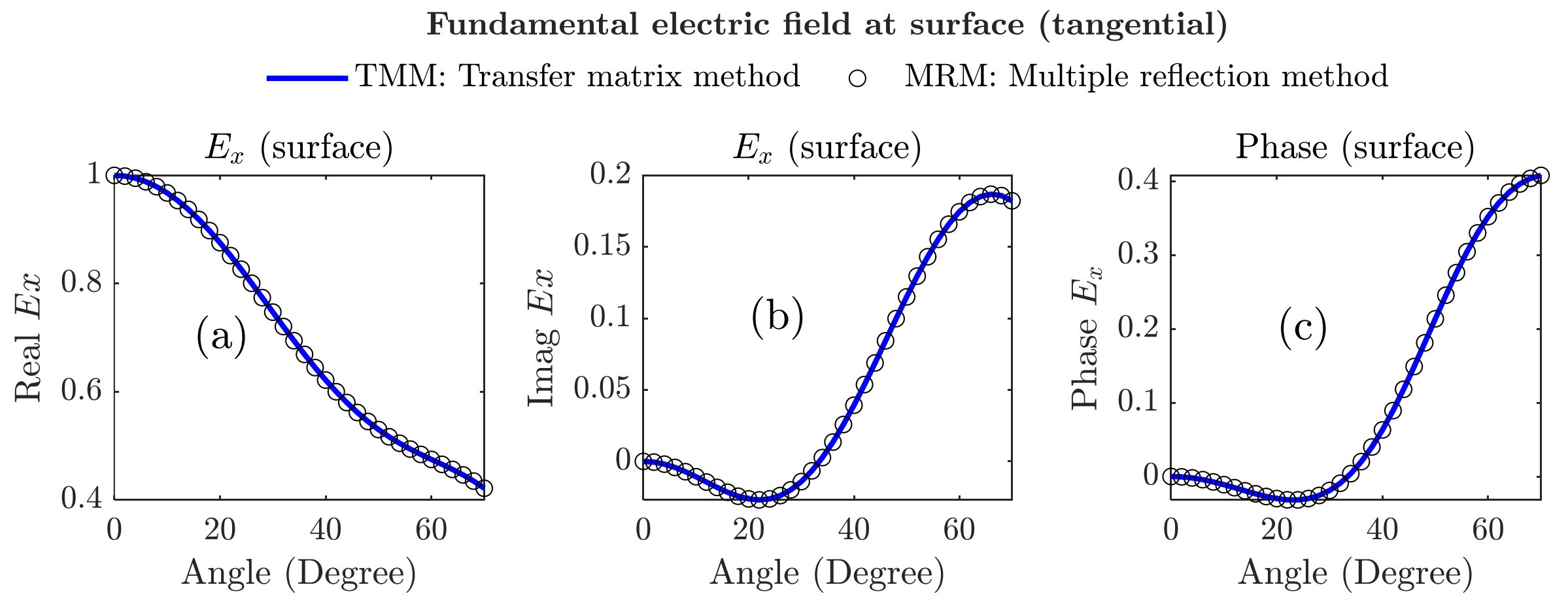

Figure 10 (a) Real (b) imaginary part of the tangential electric field component at the surface (c) phase of the field with incident angle

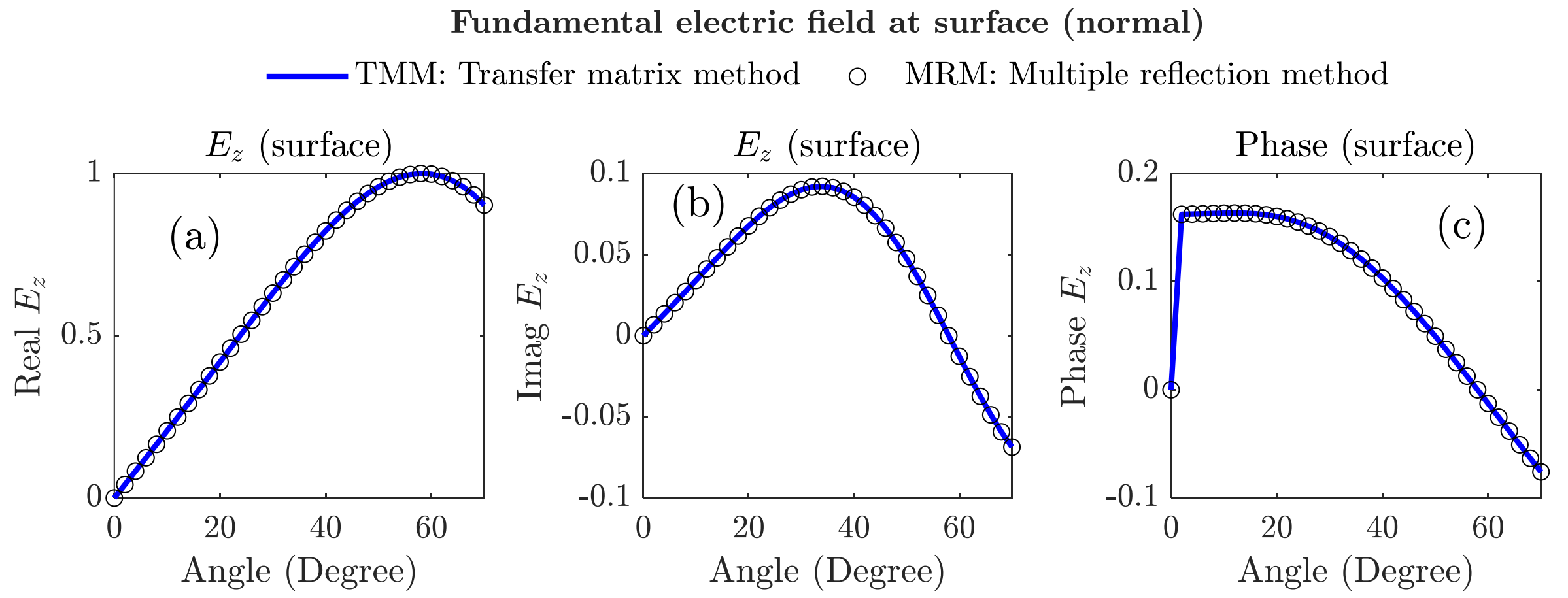

Figure 11 (a) Real (b) imaginary part of the normal electric field component at the surface (c) phase of the field with incident angle

Eq. (29) demonstrates that both normal and tangential components of the fundamental field are essential in the calculation of nonlinear source polarization. Thus, $E_x$ and $E_z$ of the fundamental wave contribute to the polarization. Hence, in contrast to the TE case, different components of polarization arise. Our choice of material enforces to generate two nonzero components of induced polarization, namely $P_x$ and $P_z$. Fig. 12 (a) and (b) illustrate the real and imaginary parts of $P_x$ at the surface (air-film) calculated from both methodologies. Identical $P_x$ ensures the identical phase evolution as depicted in Fig. 12 (c). On the other hand, equivalence of $P_z$ component at the surface is shown in Fig. 13 where the real and imaginary parts are depicted in Fig. 13 (a) and (b), while Fig. 13 (c) shows the phase variation with incident angle.

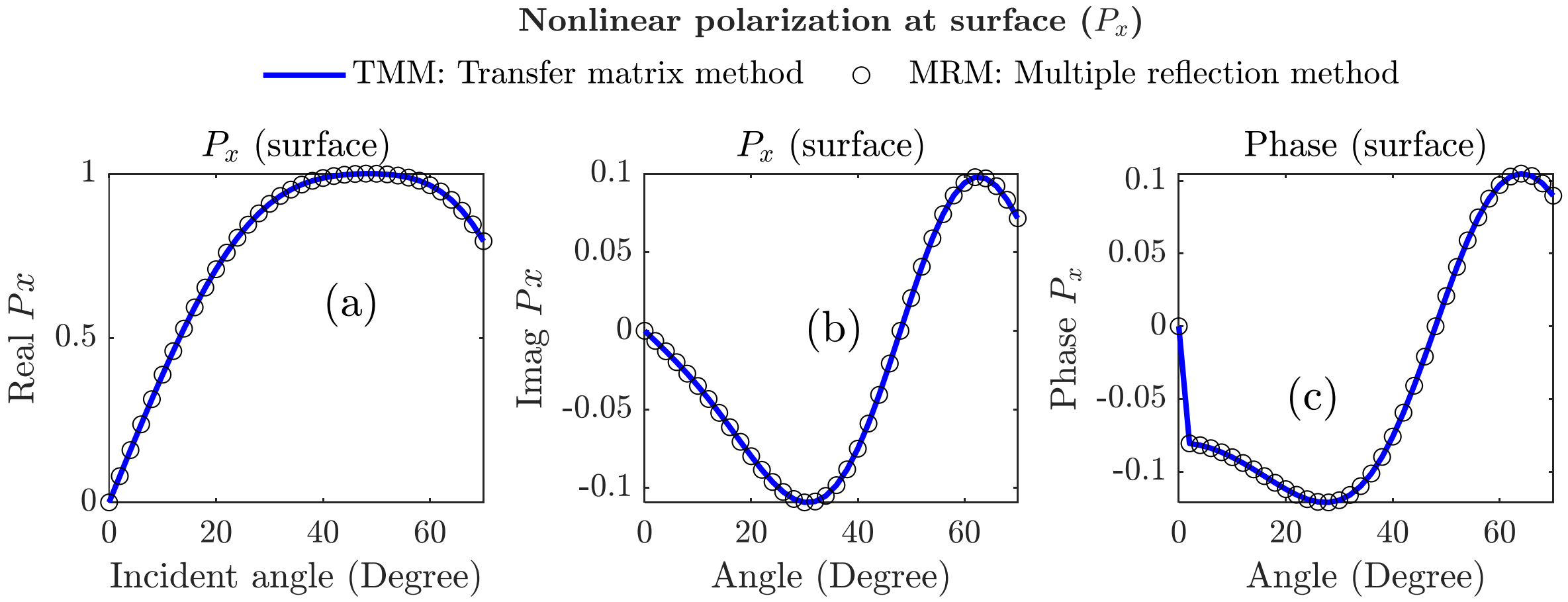

Figure 12 (a) Real (b) imaginary part of the $x$-component of the induced nonlinear polarization at the surface (c) phase of the polarization with incident angle

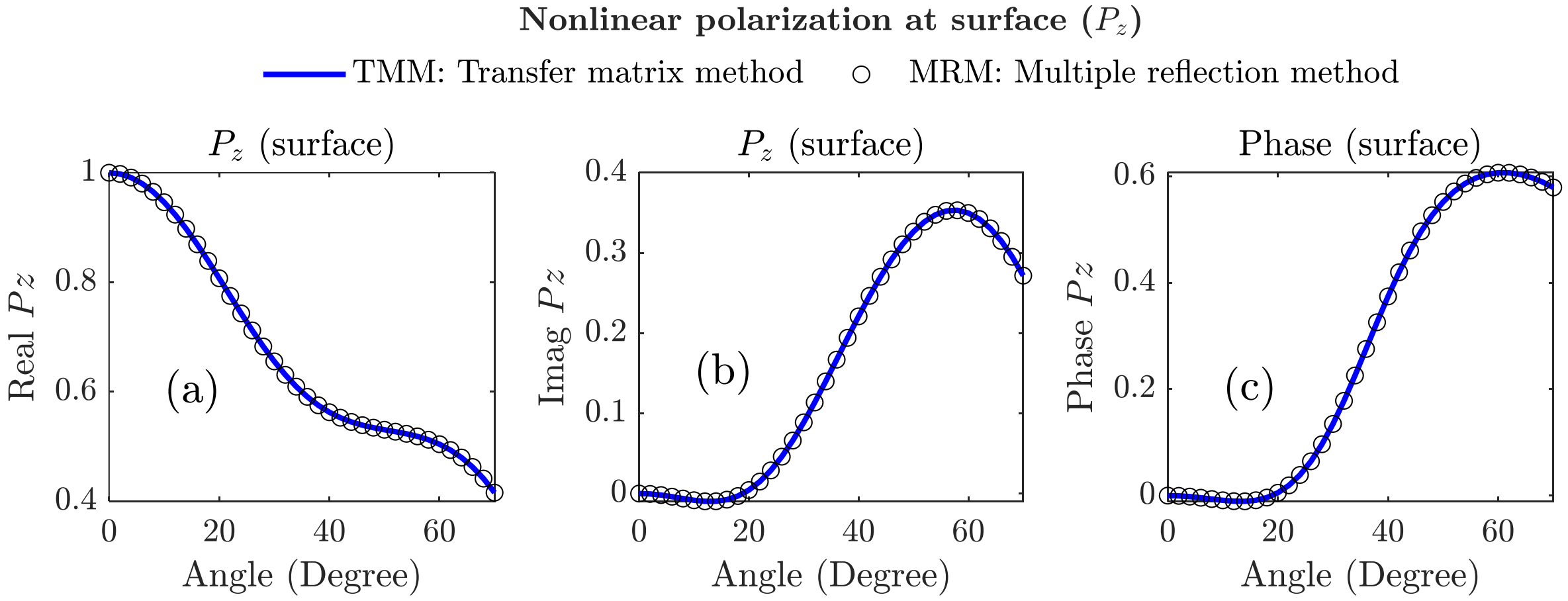

Figure 13 (a) Real (b) imaginary part of the $z$-component of the induced nonlinear polarization at the surface (c) phase of the polarization with incident angle

In contrast to the TE case, both the components of the fundamental field $E_x$ and $E_z$ are involved in the calculation of induced polarization. Consequently, phases accumulated by these electric fields must coincide to prevent discrepancies in the total induced polarization resulting from the interference of different components of induced polarization i.e. $P_x$ and $P_z$. Strict agreement between the phases of $P_x$ and $P_z$ is a prerequisite to obtain identical SHG electric field in both methods. Fig. 14 (a) and (b) verify this convergence, where the real and imaginary parts of the total surface SHG field from the thin film are shown and Fig. 14 (c) demonstrates the exact overlap of their phase evolution. This equivalence between the total surface SHG field ultimately ensures the identical total SHG intensities from both methods. As shown in Fig. 15 (a), the total intensity resulting from the interference of the intensity generated at the surface and interface of the thin film is completely identical across both methodologies. Moreover, the reflectivity of the fundamental TM-polarized wave calculated from both methods exhibits perfect agreement as demonstrated in Fig. 15 (b).

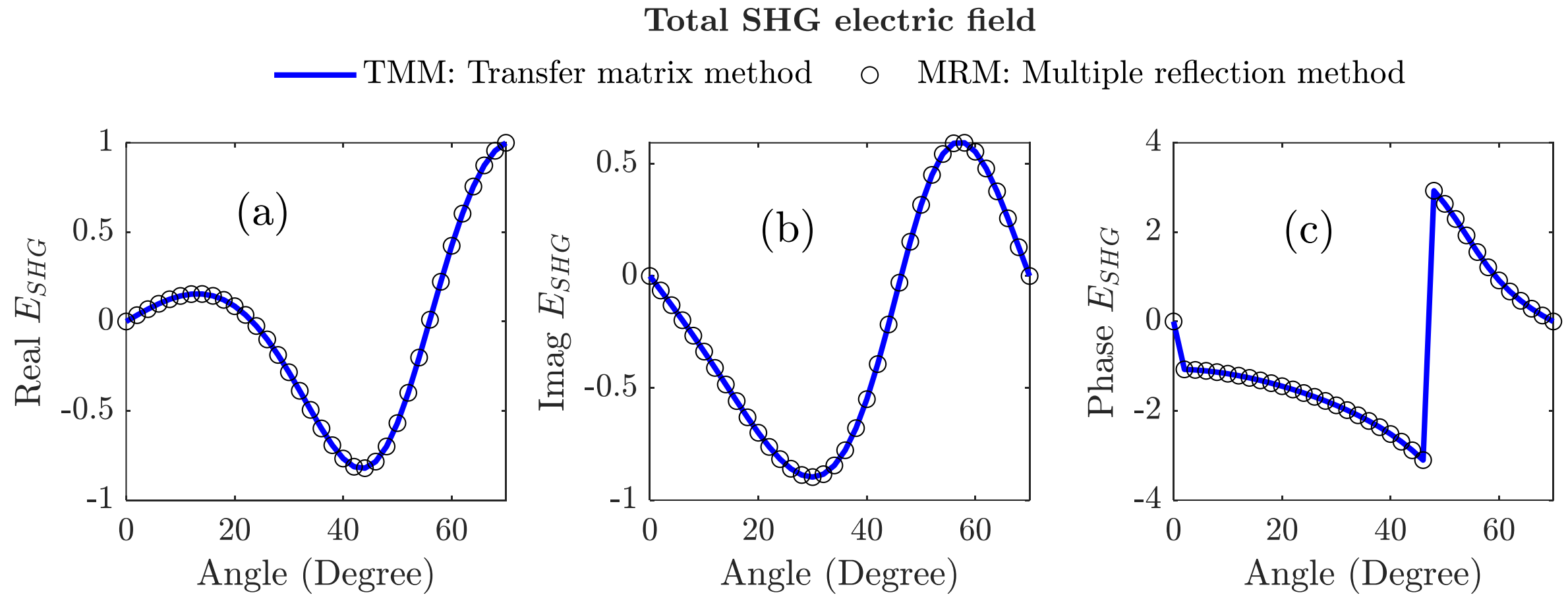

Figure. 14 (a) Real, (b) imaginary part of the total SHG electric field and (c) phase variation with incident angle

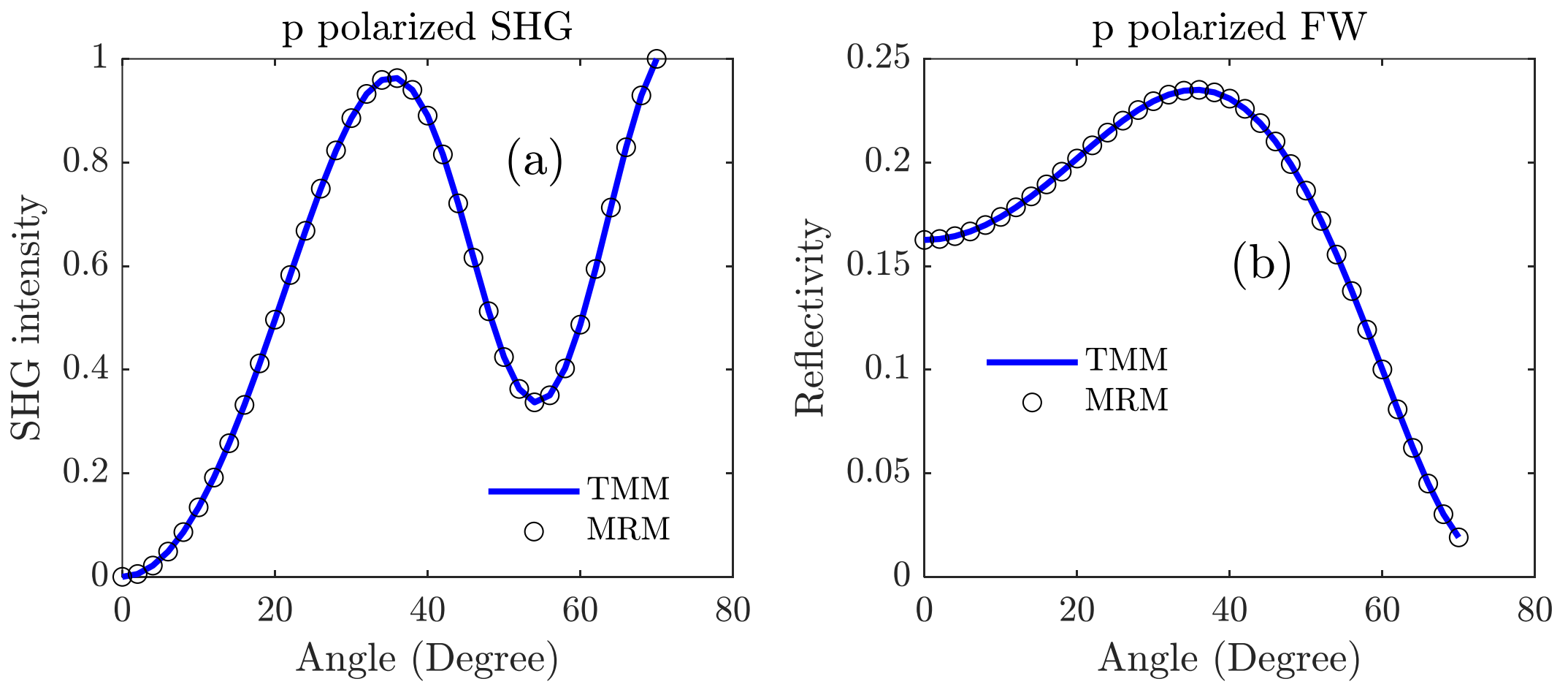

Figure 15. (a) Total SHG intensity and (b) reflectivity of the FW wave with incident angle

Thus, results discussed in Section 3.2 confirm that the multiple reflection model and TMM are equivalent methods for calculating surface-induced SHG under certain conditions. Both methods may disagree and provide distinct SHG intensity if numerical implementation is not performed accurately, as shown in Section 3.1.

## 4. Conclusion

In summary, we have established equivalence between TMM and multiple reflection modal and reported a consistent numerical implementation process for achieving identical SHG intensity. This equivalence can be obtained under certain appropriate conditions. Primarily, the position of the induced nonlinear polarization sheet is the major factor leading to ambiguity between total

SHG. Therefore, polarization sheet should be implemented consistently in both methods. Moreover, the rescaling of surface nonlinear susceptibilities in multiple reflection model and source matrix in TMM should be performed accordingly. After implementing polarization sheet consistently, multiple reflection of source field in TMM is handled according to the described procedure. As the final step, in multiple reflection model, physically acceptable projections should be considered at the total emitted SHG radiation. These three factors ensure that ambiguities in the total SHG intensity can be removed by adopting consistent numerical implementation and both mathematical frameworks can predict identical SHG intensities. Moreover, it states that surface susceptibilities $\chi^{(2)}$ inherently depends upon the convention used in the calculation. The values of $\chi^{(2)}$ calculated from one method can not be compared with another method directly and need to be reported together with the used convention.

.

**Acknowledgement**

Funded by the European Union (ERC, COINED, 101171288). Views and opinions expressed are, however, those of the author(s) only and do not necessarily reflect those of the European Union or the European Research Council Executive Agency. Neither the European Union nor the granting authority can be held responsible for them.

This work was also supported by the Czech Science Foundation (GAČR), project No. 25-17822S.

**Data availability statement**

**Conflict of interest**

The authors declare no conflicts of interest.